# Proof of enantioselectivity in a multilayer with a strong exciton polariton coupling and through asymmetric polarization

Giuseppina Simone

**Abstract**

Many important processes rely on chirality, which has led to a growing interest in improving the yield of enantioselective procedures. Here, I utilized the evanescent field created in a Kretschmann geometry to produce enantioselectivity. The plasmonic investigation involves activating a multilayer consisting of silver Ag, platinum Pt, silica $SiO_2$, and silicon Si. Hybrid waveguide and surface plasmon polariton modes define the device's plasmonic activity. The presence of micrometric features on the multilayer surface facilitates the formation of the waveguide mode. Moreover, they produce a transmitted signal that is associated with localized surface plasmon resonance. A red dye adsorbed onto a multilayer surface causes a strong coupling between excitons and polaritons. This coupling increases the radiation force and the dye-induced enhancement of radiation force supports the use of passive chirality spectroscopy to measure the optical forces acting on enantiomers. In conclusion, when a Kretschmann scheme is combined with the de-polarization, a built-in asymmetry results in a different optical flux of spectrum photons, resulting in distinct, enantioselective, and solely polarization-dependent spectral contrast, and the enantioselectivity is demonstrated for the D,L– penicillamine.

KEYWORDS: Surface plasmon polariton mode, waveguide mode, enantioselectivity, lateral optical forces, Kretschmann layout, strong exciton polariton coupling.

# 1. Introduction

Enantiomers have a wide distribution in nature and influence the functions essential for sustaining life[1]. Enantioselectivity is crucial because molecules with different handedness demonstrate distinct properties due to their interaction with the environment [2,3]. One configuration may have drawbacks while the other offers advantages in various industrial applications, particularly in the pharmaceutical industry. Enantiomers are molecules with sub-wavelength dimensions that are detectable in the UV range, but in general chiral phenomena can be weak and hard to analyze. For instance, the elastic photon's size and width make Raman scattering unsuitable for recording molecular activities like intermolecular modes between groups of molecules or the molecular modes of aromatic rings and molecular chirality [4] [5] [5,6]. When trying to distinguish between enantiomers in a solution, the separation can be enhanced with well-designed substrates [7] [8] and require a carefully selected solvent that does not induce interference with the chiral properties [9–11] [12]. Enantiomers that are left- and right-oriented might have different refractive indices, but they have opposite optical activity and then display an opposite rotation of the polarization plane as well as a phase shift between their orthogonal polarizations [13,14]. The observation of chirality in the enantiomer electromagnetic response has opened up the potential to separate handedness molecules by utilizing an optical force through passive sorting mechanisms [15,16] and the generation of lateral optical forces [15]. The lateral optical forces derive from the chiral particle interaction with the reflection of the scattered field and the transfer momentum and are exerted by an electromagnetic plane wave on chiral objects that are adsorbed on reflective surfaces. They act in a direction where the wave propagation is absent and can deflect objects with opposite helicity toward opposite directions[15]. It is important to note that achiral objects are subject to a lateral force arising from a linear momentum connected to the extraordinary spin angular momentum in an evanescent wave [17] [18]. A recent analysis has shown that lateral optical forces emerge through a direct interaction of the optical spin angular momentum with the chiral particles and, based on the particle helicity, develop toward the direction of the spin angular moment or opposite to it [19].

Here, I tested the enantioselectivity of a couple of enantiomers by using the benefits of creating an evanescent field in a Kretschmann geometry resulting in an exclusively polarization-dependent spectral contrast. It is hypothesized that the spin angular momentum creates the lateral forces necessary to produce evanescent waves with elliptically polarized transverse magnetic and electric fields for achieving chiral discrimination. To demonstrate how lateral optical forces measured on chiral molecules activate enantiomeric discrimination, a multilayer has been designed and used. A thin metal layer of Ag is deposited on top of the multilayer, with a layout that is made up of alternately low and high refractive index layers, such as Si, $SiO_2$, and Pt. I conducted the measurements using the Kretschmann geometry; this allows for coupling of the evanescent field through a prism and provides stability to the measurements by minimizing the sample susceptibility through evanescent light coupling with a focal spot [20–22].

The setup is designed to produce a hybrid plasmonic mode by combining the surface plasmon polariton (SPP) mode, created by a thin layer of Ag, with the waveguide mode, which is generated by a sequence of layers. I investigate the impact of strong coupling between the waveguide and SPP modes on metal losses of the SPP component and whether a narrow plasmonic resonance given by the hybrid can enhance molecular sensitivity[23]. During the investigation, I analyzed the impact of the micrometric features present on the surface of the multilayer. These features are created during the groove etching and coated with $SiO_2$ and Ag; I demonstrate that the features aid in the creation of the waveguide mode. Moreover, the micrometric features in combination with the local electromagnetic enhancement cause resonant plasmonic modes and scattered wavevectors. This produces vibrations at low frequencies that are associated with a localized acoustic mode and a localized surface plasmon resonance. I conducted a study on enhancing radiation force and evanescent field to impact enantiomer separation[24]. Using a two-harmonic oscillator model, I explored the strong coupling of exciton/polariton by adsorbing a red dye on a multilayer.

It is interesting to observe that the readability of the enantiomers can also be tied to the low-frequency range that characterizes this study. Modifying the symmetry of the excitation beam could significantly affect long-range interactions in the lower frequency range. Since biomolecular mechanisms belong to the long-range interaction category, they are more readily observable in the low-frequency domain. Indeed, they demonstrate greater sensitivity to alterations in the symmetry of the excitation beam in low-frequency regions where their activity is most pronounced, as opposed to high-frequency regions. Hence, I suggest that utilizing the aforementioned method to observe biological events could aid in advancing the field of biomolecular analysis.

## 2. Experimental
### 2.1 Detection of surface plasmon resonance

I used the Kretschmann configuration for the optical measurements [25–27]. The system was excited with a tunable continuous-wave laser $\lambda = 650$ nm via a beam splitter. For each angle, the system was given the time to reach a steady state before measuring the signal. For the measurements, the laser was collimated (Thorlabs RC12FC-P01) and *p*-polarized by a double Glan-Taylor Calcite Polarizer; the noise was lowered with an optical iris diaphragm (Thorlabs, D25SZ). The reflected beam was collimated and filtered (Thorlabs, Glan Thomson polarizer, 650-1050 nm) before being collected by Si photodiode ($\lambda_0 = 960$ nm bandwidth) to gather the signal; besides, a ceramic disk capacitor was used to cut the noise[28,29].

### 2.2 Numerical model

The code for the theoretical simulation was written in Matlab (Ed. 2023). The script implemented a frequency-domain modal method known as the Rigorous Coupled Wave Analysis RCWA in agreement with the model presented in ref. [30] and it computed the diffraction efficiencies and the diffracted amplitudes of finite-size structures composed of the stack layer. RCWA relied on the computation of the eigenmodes in all layers of the structure with the micrometric features based on a Fourier basis [31] and on a scattering matrix approach, for recursively relating the mode amplitudes in the different layers. The sum of the reflected and transmitted efficiency and the loss per layer divided by half period were used to determine the absolute error ($<10^{-5}$). The refractive index for each layer was taken from the dedicated source: the fused silica refractive index was taken from the WVASE library, Ag and Pt, Si and $SiO_2$ from Palik[32] NBK7 glass from Schott (Table 1). The wavelength and the angle of incidence $\vartheta$ were used as sweeping parameters during the analysis.

### 2.3 Coupling Model

The dispersion of the hybridized mode was achieved by solving the eigenvalue problem of the two-coupled oscillator model represented by the Hamiltonian of $2 \times 2$ matrix:

$$H = \begin{pmatrix} E_{spp} - \dfrac{i\gamma_{spp}(\vartheta)}{2} & \dfrac{\hbar\Omega}{2} \\ \dfrac{\hbar\Omega}{2} & E_m - \dfrac{i\gamma_m}{2} \end{pmatrix}$$

where $E_{spp}$ is the energy of the plain cavity, $E_m$ is the energy of the excitons of rhodamine G6, and the line widths are $\gamma_m$ and $\gamma_{spp}$, this latter changes with the angle according to the behavior of the angular dispersion.

**Table 1.** Refractive index and geometry of the layers in multilayer and micrometric features

| multilayer | | Micrometric feature | | Refractive index |
|---|---|---|---|---|
| layer ID | height | parameter | size | $n_i$ @ λ=650 nm |
| Pt | 500 nm | gap | ~1 µm | Pt  2.38+4.26i |
| $SiO_2$ backside | 500 nm | diameter | 1.4 µm | $SiO_2$  1.44 |
| Si | 30 µm | Si height | 30 µm | Si  3.85+0.018 i |
| $SiO_2$ frontside | 1 µm | $SiO_2$ height | 1 µm | Ag  0.052+4.41i |
| Ag | 30 nm | Ag height | 30 nm | rG6 1.35+0.0005i |

## 3. Results and Discussion
### 3.1 Rationale of design and characterization of the multilayer

Figure 1a shows the cross-sectional schematic of the structure used in this study, the multilayer and groove are labeled by the dashed rectangles. The multilayer is made up of alternating low and high refractive index materials, including Si, $SiO_2$, and Pt, with a thin layer of Ag on top. The properties and geometric size of the multilayers are summarized in Table

1. For the analysis, micrometric features that are shown in Figure 1b decorate the surface of the multilayer and they are represented as columns having an average diameter and height $d = 1.4 \pm 0.3\ \mu m$ and $h = 30 \pm 5\ \mu m$ and a gap $b = 1\ \mu m$.

I used a Kretschmann-based coupler in an angle-resolved frame and monochromatic *p*-polarized light to measure the surface wave and match the in-plane wavevectors in the stop region of the multilayer. The angular reflectance spectrum (Figure 1c) displays a drop (1) at an incidence angle of $\vartheta = 43.48$ deg and a second minor dip (2) at $\vartheta = 41$ deg; an analytical prediction of a *sans*-feature multilayer suggests that the reflectivity drop should occur at $\vartheta = 41$ deg. Therefore, when comparing the two spectra, it becomes evident that the dip (2) is representative of the surface plasmon polariton mode identifiable as a Bloch surface wave [33], while the plain multilayer forms the drop (1). I hypothesize that a TM waveguide-mode can be identified as mode (1) at the interface of the multilayer Pt/SiO$_2$/Si/SiO$_2$ and the Ag, and the micrometric features enhance this behavior to create a distinct signature for the plasmonic system; a slot waveguide mode develops between two silicon micrometric features. Mode (2), on the other hand, occurs at the Ag/air interface. Figure 2a offers a schematic of the modes. Usually, the waveguide modes do not need a coupler for excitation; nevertheless, if the hypothesis of the waveguide mode excitation is validated, the Kretschmann setup including a multilayer is able to generate a hybrid waveguide-SPP mode [34][35]. The hybrid mode will offer the advantage to be excited in *p*- and *s*-polarization, and it can be valuable for molecular studies. In addition, both the *p*-polarized hybrid modes of *p*-waveguide and *p*-SPP can be produced on a same metal layer. These modes are stimulated at different interfaces of the same metal layer, which may result in the manifestation of the anti-crossing effect in their dispersion curves.

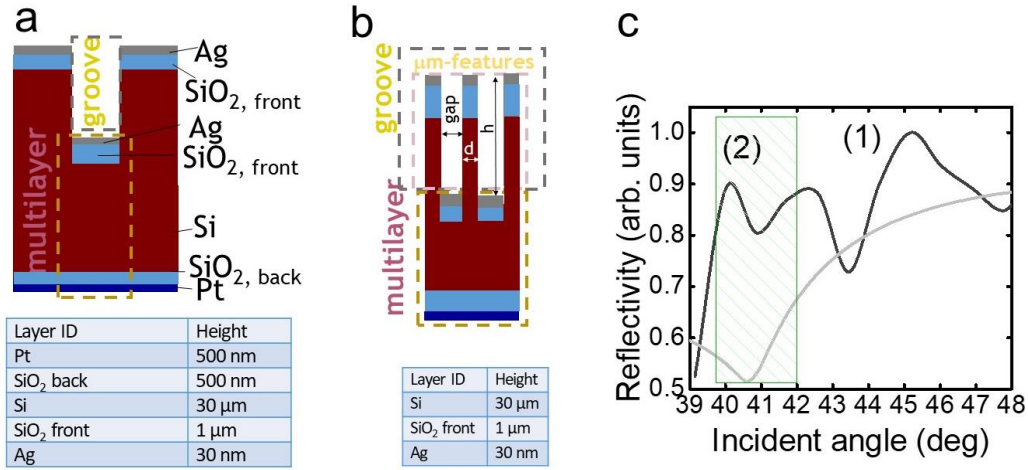

**Figure 1.** Multilayer and micrometric features (a) Out-of-scale cross-section of groove and multilayer. Dashed rectangles isolate the groove and the multilayer. Bottom table: sizes of the multilayer. (b) Out-of-scale scheme of the multilayer decorated by the micrometric features. Bottom table: sizes of the micrometric features. (c) Reflectivity measured at TM polarization in air compared with an analytical prediction (Winspall) relative to a plain multilayer. Symbol: modes, (1) SPP, (2) multilayer.

My hypothesis has been validated through an analysis of the electromagnetic field. The study was conducted using the geometry illustrated in Figure 2a, which included the multilayer and micrometric features. A transmission medium, such as air, surrounds the sample according to the prism coupling shown in the inset of Figure 2a, and the laser traverses the prism before coupling to the multilayer. For the geometry described in Table 1, the electromagnetic field is shown in Figure 2b. Upon analyzing $E_x$ an enhancement was observed at the Ag/SiO$_2$ interface (panel (i)). This finding supports the hypothesis of a mode generated between the multilayer and the Ag; the latter can be identified as the TM waveguide mode. The analysis of $E_y$ displays a comparable scenario that arises at the top of the features (panel (ii)).

Multiple geometries with different behaviors can be studied by adjusting parameters such as height, gap, and distribution of micrometric features. However, the analysis showed that the gap is the most influential factor for tuning the electromagnetic behavior; in particular, the *x* and *y* components of the field achieve the maximal value for a gap $b \approx$ 1 $\mu m$ (Figure 2c). Furthermore, experiments have been conducted under the different angular apertures of the incident angle where the electric field strength demonstrates a peak at the resonance, as seen in Figure 2c. The nature of the two modes was also studied by recording the reflectivity according to the angular aperture of the incident beam within a specific wavelength range. Notably, when examining the reflection curves at $\vartheta = 43.5$ deg, two modes were identified (Figure 3a). The first mode occurs at a wavelength of 550 nm and has a broader spectral range and lower energy. This mode is not influenced by changes in the angle, and it results in a small mode volume and then is less relevant for the implementation of sensitive devices. Due to its behavior, it is classified as the SPP mode. In addition, two modes with peaks at wavelengths of 475 nm and 620 nm, respectively, flatten out of resonance. Unlike the SPP mode, these modes have a slight linewidth and a significant shift and are classifiable as waveguide modes[36]. Figure 3a provides a dispersion diagram and shows the hybridization of SPP and waveguide modes. The hybrid waveguide-SPP plasmonic mode dispersion relation between the two components is affected by the repulsive forces of these dispersion curves; the SPP resonance losses are decreased, and the spectral width is thinner than the SPR resonance. This anti-crossing effect suggests the existence of a hybrid waveguide-SPP mode, in which the waveguide state exchanges energy with the SPP state and establishes a strong coupling regime.

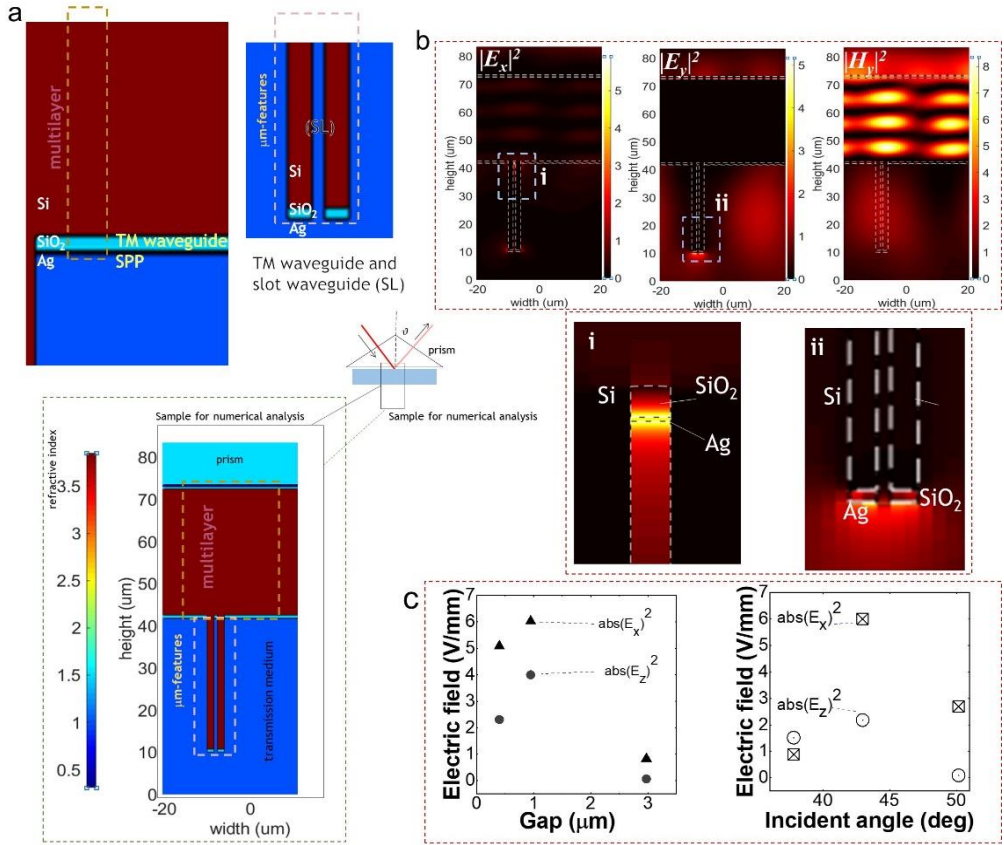

**Figure 2.** Mode characterization. (a) Top: Hypothesis of the existing modes. Bottom: Refractive index map of the modeled layout. The dashed rectangles isolate the fundamental domains. Color bar: refractive index. (b) Top: numerical predicted electric field and magnetic fields at $\vartheta = 43.48$ deg and TM polarization. Bottom: zoom at the bottom Ag/SiO$_2$ interface (i) and at the top of the columns (ii). Color bar: field intensity. (c) Top: electric field-gap and, bottom: electric field -incident angle trends.

Additionally, the absence of waveguide modes in the *sans*-features multilayer dispersion diagram highlights that the microstructure plays a crucial role in promoting the modes that arise at the interface of SiO$_2$ and Ag and therefore, these modes serve as a distinctive signature of the microstructure itself.

### 3.2 Characterizing the vibrational regime

In the previous section, it was shown that micrometric features have an impact on the hybridization of two modes [37]. To analyze their influence on the optomechanical response, I introduced an optical power of 5.0 mW into the system at an incident angle of $\vartheta = 43.48$ deg to measure the scattering of the micrometric roughness. The photodiode's output was monitored using an oscilloscope and displayed in the frequency domain by a spectrum analyzer. The Fast Fourier Transform FFT spectrum shows three fundamental peaks at the resonance, consisting of a primary peak centered at $\omega_m = 96\ MHz$ and two secondary peaks at $\omega_m \pm 2$ MHz (Figure 3b). When comparing the spectra recorded on and off-resonance, it can be observed that the intensity of the side peaks changes depending on the angle of incidence, with the on-resonance peak being the highest. However, there is no frequency shift between the on-resonance and off-resonance spectra, indicating that the phenomenon is not caused by a wavevector effect. Instead, it is due to acoustic vibrations that are confined to the roughness. By using the relationship $f = v/d$ ($v$ that is the sound propagation speed in metal $v = 3 \times 10^3$ m s$^{-1}$ and $d$, the average roughness), the acoustic frequency vibration can be estimated and the calculation results in an analytical value of $f = 100\ MHz$. The fact that the experimental results are similar to the calculated value supports the assumption about the nature of the transmitted signal.

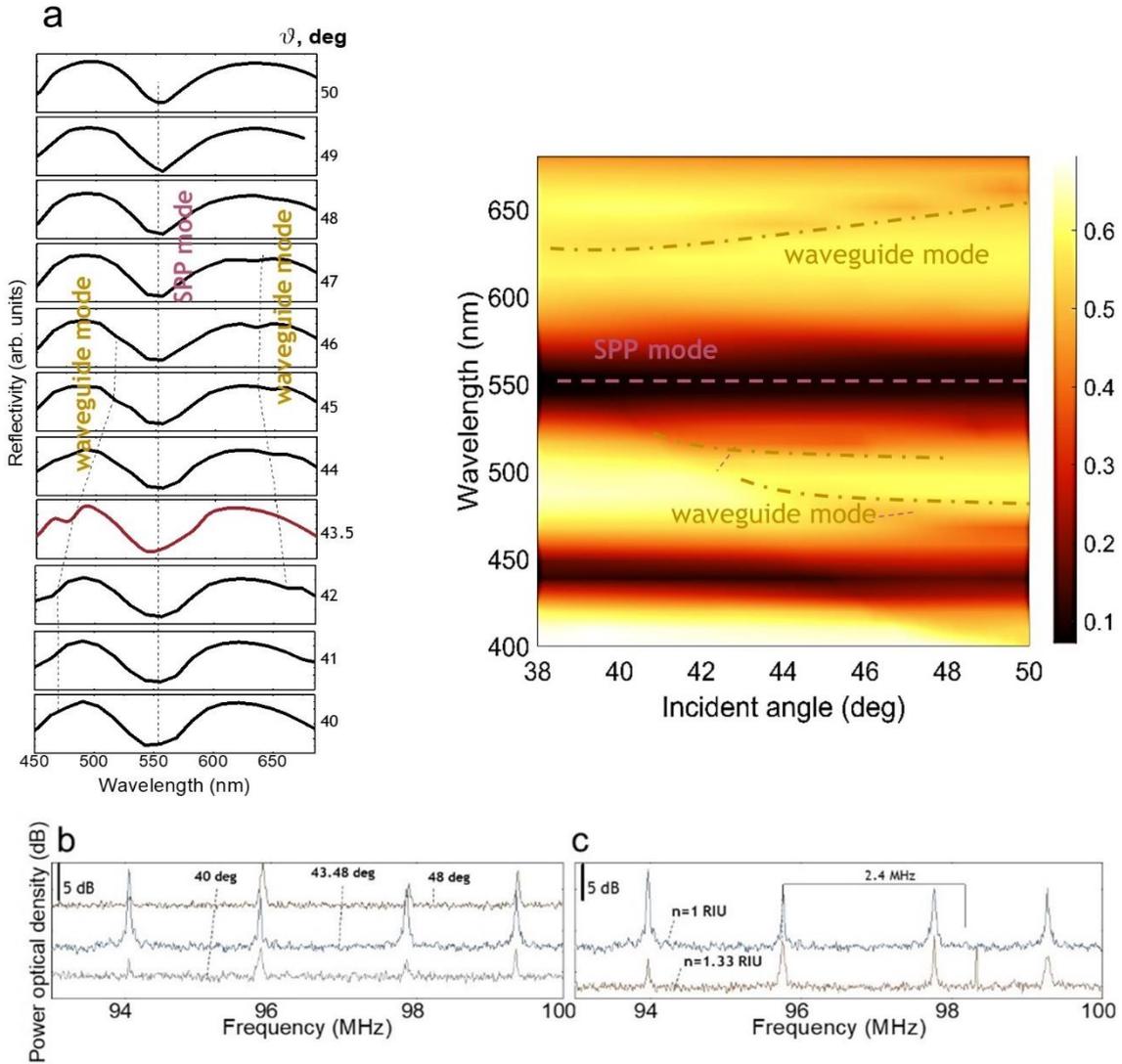

**Figure 3.** (a) Reflectivity according to the incident angle; dashed lines highlight the mode dips. Right: Dispersion of the multilayer. The dashed lines underline the modes. Color bar: reflectivity. (b) FFT spectrum of the optomechanical hybrid system at several angles of the angle of incidence. (c) FTT spectrum of the optomechanical hybrid system at $\vartheta = 43.48\ deg$ in an aqueous environment ($\vartheta_{H2O,res} = 46.5\ deg$).

The measurement was repeated in water with the angle of incidence fixed at the measured angle of resonance $\vartheta_{h2o,res} = 46.5$ deg. When the surrounding medium air is replaced by water ($n_{H2O}$=1.33 RIU against $n_{air}$=1.0 RIU), a clear dependence on the refractive index is measured (Figure 3c); indeed, the right peak $\omega > \omega_m$ in water produces a doublet resonance splitting and higher frequency shift ($\Delta\omega = 2.4\ MHz$). The influence of the refractive index of the surrounding medium, together with the observed plasmonic enhancement and the wavevector-independent nature of the frequency evidence that the transmitted signal is related to an intermediate state near the rough surface. Furthermore, the dependence on the surrounding medium refractive index indicates a correlation with a localized resonance phenomenon, and it is plausible that the intermediate state is not confined at the surface, but it extends in free space where the surface plasmon polaritons travel. On the side of the surrounding medium, I have estimated that the spatial extent of the layer penetration depth is $\delta = 5.04\ nm$ (Supporting Information S1). Since the penetration depth is thinner than the exciting beam wavelength $\lambda_{exc} > \delta$ [38], the observed spectra are attributed to localized surface plasmons. The assessment for determining the influence of the localized plasmonic resonance on the transmitted signal is beneficial for the sensitive analysis and will offer valuable insights that can aid in making well-informed decisions and drawing sound conclusions.

In order to measure the enantioselectivity of the multilayer and maximize the performance through the enhancement of the radiation rate, rhodamine G6, a fluorescent dye, was adsorbed onto the surface. This creates two mechanical oscillators in the system: the multilayer polaritons and the rhodamine exciton, which amplifies the vibrational signal [39–41]. The absorption and emission of the dye were measured at wavelengths of $\lambda = 561\ nm$ and $\lambda = 571\ nm$, as shown in Figure 4a. To examine the interaction between the exciton and multilayer, different concentrations of the dye were tested in an angular frame. As the concentration increased (Figure 4b), the SPP mode flattened while the depth of waveguide mode intensified. The strong coupling between the multilayer and the exciton causes an increase of losses, resulting in a decrease in the intensity of $E_x$, as seen in Figure 4c. This leads to a contraction of the frequency shift in waveguide modes, resulting in an increase in mode volume, and leads to the formation of both a high and a low hybrid polariton (LP and HP) as shown in the dispersion diagram (Figure 4d). Next, I examined the system's angle-resolved reflectivity (Figure 4e) [42][43]; both upper and lower polariton branches can be observed as local minima and are split around the exciton and polariton dip. When the energy of these modes is plotted against the incident angle, the dispersion diagram in Figure 4g is achieved and a model with two coupled oscillators is used to fit the results. The model that has been presented in Experimental accounts for the plasmon and exciton bare energies ($E_{spp}, E_m$), as well as the strength of the exciton/polariton coupling and the intrinsic radiative rate of the rhodamine $\gamma_m$, and the cavity radiation rate $\gamma_{spp}(\vartheta)$ (Figure 4f and Table 2). Based on the findings, it has been highlighted that when the angle of incidence is at $\vartheta = 46.5$ deg, the anti-crossing energy is observed to be $\hbar\Omega = 440$ meV. This energy is measured as the distance between the hybridized modes at the point where the exciton and the polariton (dashed lines in Figure 4g) intersect each other.

Moreover, the agreement between the experimental and analytical fit enables us to extract significant parameters. For instance, based on the model, there exist two distinct solutions HP and LP, and the plexciton states involve anti-crossed bands. At $E_{spp} = E_m$, the harmonic model estimates the anti-crossing energy as $\hbar\Omega = \sqrt{4G^2 - \frac{\gamma_{spp}(\vartheta) - \gamma_m}{4}}$ that is a function of the coupling strength $G$ and the radiative rates, and it provides a validation condition for the strong coupling between the exciton and the polariton, as given by a hierarchy between the anti-crossing energy and the radiation rates, $\hbar\Omega > \frac{(\gamma_{spp} - \gamma_m)}{2}$. Since the dissipation energy of the hybrid cavity can be calculated to be $\gamma_{spp}(\vartheta) = 8$ meV, and the radiation rate of the rhodamine to be $\gamma_m = 0.15$ meV [44], for the studied multilayer, I verified the hierarchy $\hbar\Omega > \frac{(\gamma_{spp} - \gamma_m)}{2}$, which proves the strong exciton/polariton coupling and estimates the coupling strength to be $G = 220$.

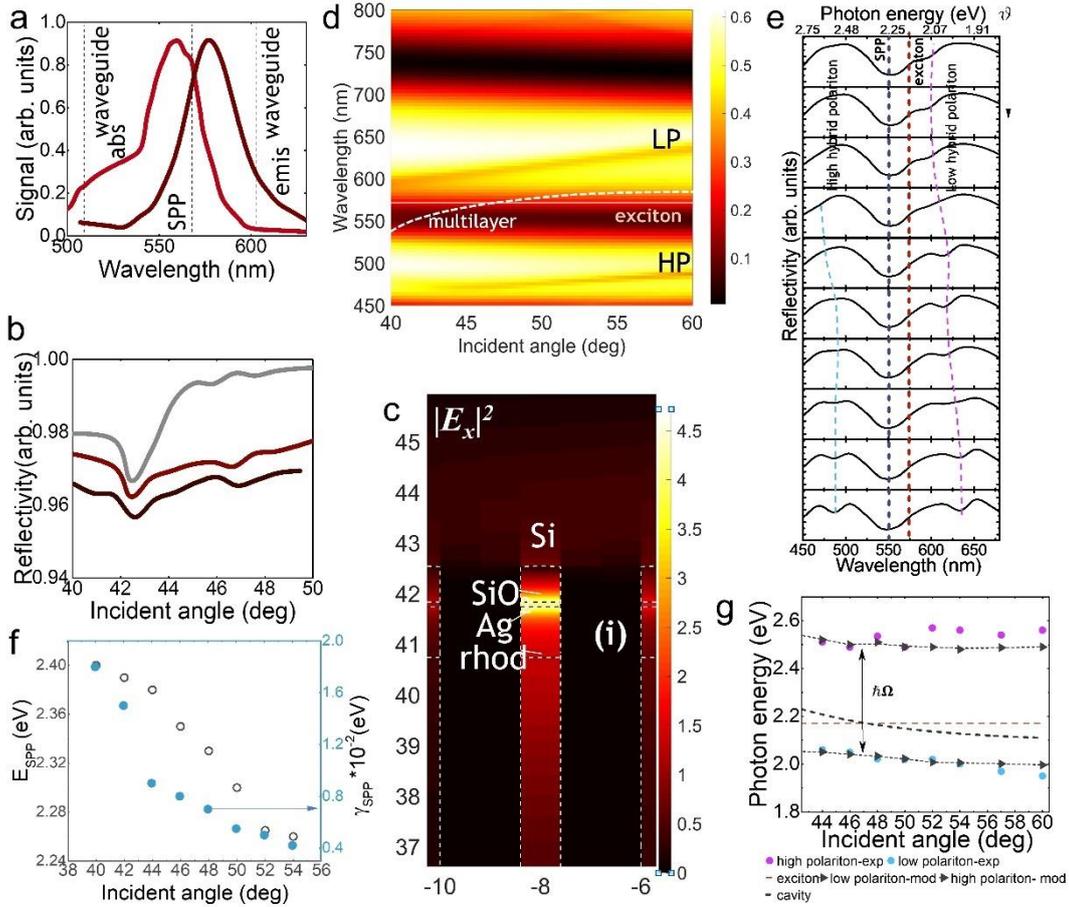

**Figure 4.** Exciton/polariton coupling. (a) Left: rhodamine absorption/emission; dashed lines evidence the multilayer modes. (b) Experimental reflectance spectra about different concentrations of rhodamine (from bottom to top: 5.8 mM, 0.34 mM, 0.02 mM) at $\lambda = 650\ nm$. (c) Electric field $E_x$ after dye adsorption at the gap (i). (d) Angular dispersion of the multilayer/dye. Color bar: reflectivity. (e) Experimental *p*-polarized reflectivity at different angles from 40 to 50 deg. Dashed lines: hybrid upper polariton and lower polariton branches. (f) Angular $E_{spp}$ and $\gamma_{spp}$. (g) Analytical (solid lines) and experimental (dots) energy of the exciton/polariton system.

In addition, I found that the radiative rate outperforms the cavity decay of a factor $\frac{\gamma_f}{\kappa^*} = 14.4$ (Supporting Information S2). It is also interesting to observe that the results indicate a system that can detect both high and low-speed events by combining a high-frequency mode in the terahertz range emitted by molecules with a low-frequency mode in the megahertz range from the cavity. This could potentially improve molecular sensing, which currently has limitations due to its properties[45].

**Table 2.** Exciton/polariton factors and their value

| multilayer (polariton) | | |
|---|---|---|
| angle, deg | $E_{spp}, eV$ | $\gamma_{spp}, eV$ |
| 40 | 2.4 | 0.018 |
| 42 | 2.39 | 0.015 |
| 44 | 2.38 | 0.009 |
| 46 | 2.35 | 0.008 |
| 48 | 2.33 | 0.007 |
| 50 | 2.3 | 0.0055 |
| 52 | 2.265 | 0.005 |
| 54 | 2.26 | 0.0042 |
| **rhodamineG6 (exciton)** | **exciton/polariton** | |
| $E_m = 2.5\ eV$ | $\hbar\Omega = 440\ meV$ | |
| $\gamma_m = 0.15\ meV$ (*) | | |
8888

(*) For estimating $\gamma_m$ the intrinsic radiative decay rate, I refer to the Quantum Yield QY and the lifetime τ ($\gamma_m = QY/\tau$) of the rhodamine as reported in ref. [46].

### 3.3 Demonstrating enantioselectivity

I exploited then the advantages of the strong exciton/polariton coupling for fostering the enantioselectivity based on the influence of the radiation rate enhancement on the evanescent field. I recorded the signals of D- (right) and L- (left) penicillamine, which were diluted in ultra-pure water with a concentration of 1 mM and 10 mM, as well as the racemic solutions. The samples were excited with a *p*-polarized laser beam; I observed that when in contact with the L-penicillamine, the *p*-polarized beam causes a splitting of the mode recorded at 90 $MHz$ (Supporting information, S3). This phenomenon was also shown by the racemic spectra and can be used as a distinguishing feature of the sinistrorsal molecule. The mode splitting supports the hypothesis that there is a discernible difference in the interaction and scattering process of light with the D- and L- enantiomer, allowing for discrimination between the two molecules through lateral forces [19,47]. In Figure 5a(i), I explain how the process works; the electromagnetic wave with helical polarization creates moments in the chiral object. The strength of these moments depends on how well the helicity of the electromagnetic wave matches the molecule chirality. As a result, the chiral objects produce radiation with a helicity direction that depends on the enantiomers and generates the sorting forces. To utilize the sorting principle based on the induced moment, the optical setup was adjusted as shown in Figure 5a(ii). This involved using a circularly polarized wave to excite the samples during total internal reflection, to maximize the optical spin angular momentum density[15]. The direction of spin angular moment density can be approximated by considering the helical polarization of the electromagnetic field. Besides, the spin angular moment generates an evanescent field that can interact with the sample positioned on the resonator's surface.

During the experiment, a second prism was paired with the first prism to capture the evanescent wave using a frustrated total internal reflection mechanism [48]. The resulting output was filtered to isolate the *s*- and *p*- polarized signals and then sent to a photodetector, which displayed the power as a function of frequency. In this configuration, the photon tunneling allowed the signal to be used for the analysis of molecular rotation by converting the generated quanta of the polaritons to photons via a dynamical Casimir effect which overcomes the limitation of discrimination between the pairs.

I analyzed the L- and D-stereoisomers of penicillamine at two different concentrations (1 mM and 10 mM). In Figure 5b, I show the spectra within the frequency range of $\omega = 87 - 94\ MHz$, while Figure 5c zoomed in on the peak at the frequency of $\omega = 89.6\ MHz$, which is the typical frequency of the systems including the enantiomers according to spectral analysis. It is worth mentioning that, just like in the water experiment shown in Figure 3c, the enantiomers affect the refractive index, and lead to a frequency shift in the spectra with respect to the multilayer. Therefore, the frequency shift observed in the spectra is evidence that the inelastic scattering is confined to the roughness and not due to the wavevector effect.

Besides, the inherent selectivity of the system for the two penicillamine enantiomers can be confirmed by using a substrate that has a weak coupling between the exciton and polariton (Supporting information, S3). However, the findings that I show in Figure 5c display a greater yield of enantioselectivity can be achieved when using a substrate that exhibits a strong exciton/polariton coupling during measurements.

Based on the spectra, the interaction and scattering of light with D- and L- molecules create forces that vary in strength depending on the molecular polarizability. In particular, the L- molecule better interacts with the p-polarized beam, as shown by the schematic in Figure 5a and by the plot in Figure 5b. These forces can discriminate between the two specular molecules[40].

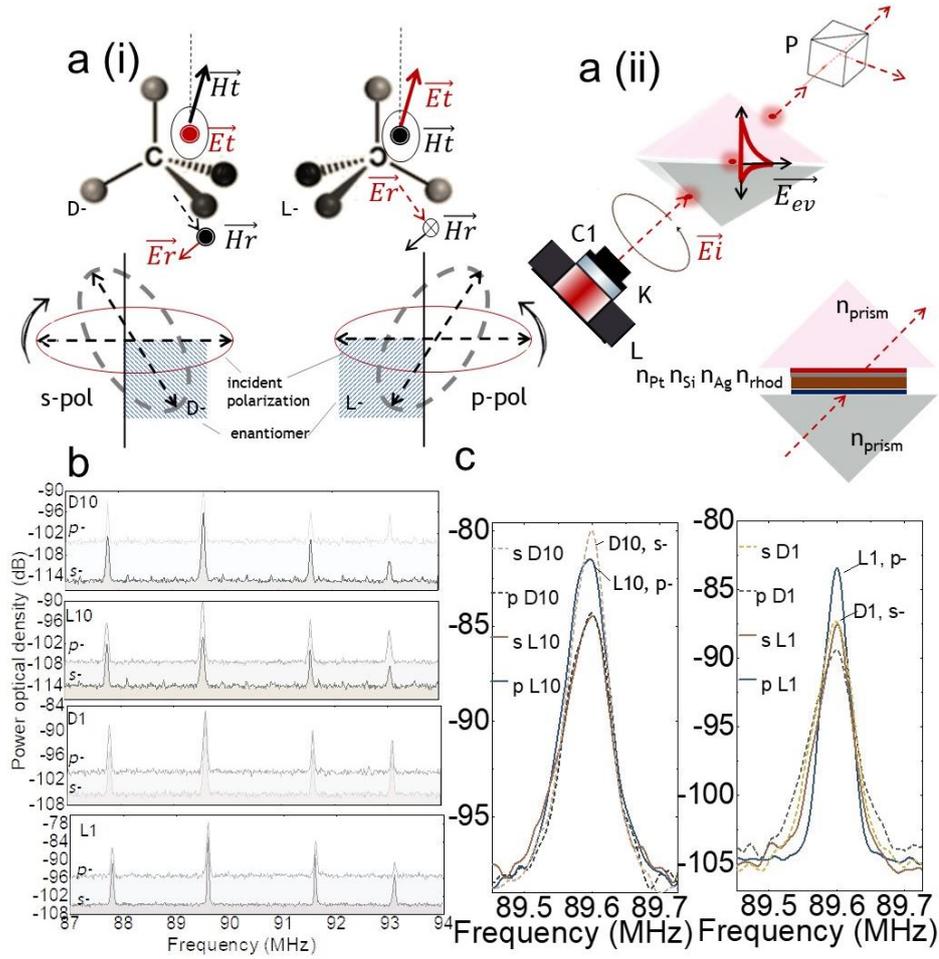

**Figure 5.** Passive chirality spectroscopy. (a): (i) Working principle of enantiomer discrimination. $X_i$, $X_r$, $X_t$: incident, reflective, and transmitted electric (X=E) and magnetic (X=H) fields. (ii) Setup for tracking the chirality; P: Glan-Taylor Calcite Polarizer; C: condensator; K: iris diaphragm; L: laser. $\vec{E_i}$ and $\vec{E_{ev}}$ incident and evanescent electric field. Up inset: top view of the prism coupling for collecting the evanescent wave. Bottom inset: prisms-sample arrangement. (b) Transmission spectra L1, D1: 1 mM, L10, D10: 10 mM. (c) Discriminative peak at 89.6 $MHz$; left: 10 mM, right: 1 mM.

### 4. Conclusions and outlook

The plasmonic resonant behavior of a multilayer with a rough surface was comprehensively studied. The surface topology is characterized by having certain micrometric features that have been demonstrated to affect the system's optical response. It is worth noting that when it comes to achieving a low-cost, highly controllable, and flexible process, cleanroom technologies and the process-dependent formation of micrometric features can be limiting factors in this investigation. The multilayer plasmonic activity relies on the hybrid mode generated by the waveguide and surface plasmon polariton modes observed in the system. Additionally, when combined with the local electromagnetic enhancement, the micrometric features result in resonant plasmonic modes, scattered wavevectors, and low-frequency vibrations. These vibrations are attributed to a localized acoustic vibration linked to a localized surface plasmon resonance. An exciton/polariton strong coupling has been discussed with the adsorption of rhodamine on the surface of the multilayer decorated by micrometric features, which has been documented to be able to increase the radiation force. The enhancement of the radiation led by the rhodamine enhances the lateral optical forces fundamental for the activation of a mechanism of passive chirality spectroscopy. Based on this principle, enantioselectivity has been demonstrated for the D,L– penicillamine.

Conventional methods fail to differentiate between D- and L- configurations without the aid of specific substrates or sample preparation techniques. Interestingly, the readability of the enantiomers can be linked to the low-frequency range of the multilayer being studied. At low frequencies, long-range interactions are more susceptible to changes in the symmetry of the excitation beam, and since biomolecular mechanisms belong to this category, they are more easily observed in the range here analyzed. This is also where their activity is most pronounced, and they exhibit greater sensitivity to changes in the symmetry of the excitation beam. Therefore, I suggest that using this method to observe biological events could advance the field of biomolecular analysis.

**Conflicts of interest**

There are no conflicts to declare.

# Supporting information

## §S1. Penetration depth

For the penetration depth in air, only the Ag layer is taken into consideration, and the geometry is illustrated in Scheme S1. The penetration depth of the evanescent field can be correctly estimated by the approximation of $1/k_y$, in which $k_y$ is the wavevector perpendicular to the interface.

From the Snell law, the expression of $k_y$ is: $k_2 = n_{Ag}^2 \left(\frac{2\pi}{\lambda}\right)^2 \left(\frac{n_{Air}^2}{n_{Ag}^2} - \sin^2\vartheta_{inc}\right)$, it follows that by replacing the numerical values available in Table S1, the penetration depth in air is 5.04 nm.

**Table S1.** Numerical values and parameters associated with the penetration depth estimation.

| Parameter | Height | ni @ $\lambda_{exc}$=633 nm |
|---|---|---|
| Air | | 1 |
| Ag | 30 nm | 0.052+4.41*i |
| $\lambda_{exc}$ | 633 nm | |
| $\vartheta inc$ | 43.5 deg | |
| $k_y$ | 0.2 nm$^{-1}$ | |

**Scheme S1.** Geometry and parameters assumed for the calculation of the penetration depth in air.

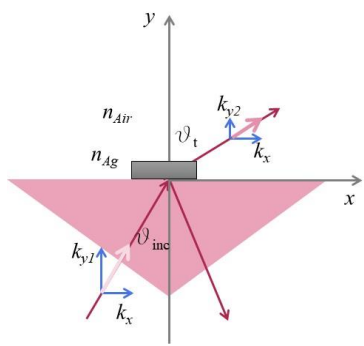

## §S2. Radiation rate

Rhodamine G6, a fluorescent dye that may make an excellent molecular emitter, was adsorbed as a radiative antenna onto the surface of the Ag layer. The new optomechanics is based on a molecular spring-mass oscillator that connects to one end of the multilayer mirror. A pump laser causes the molecules to transition from a ground to an excited state. At an emitting frequency $\omega_m$, the dipole shows an intrinsic radiative decay rate $\gamma_m$ influencing the behavior of the simple optomechanical multilayer and a negligible non-radiative decay rate. The multilayer, in turn, has a plasmonic loss $\gamma_{spp}$ at frequency $\omega_{spp}$ and, because of the crystalline Ag, loses a part of its energy to heat at a rate $\gamma_{ohm}$. Through the molecular dipole and geometrical factors, the decay rate modifies as $\kappa^* = \frac{V_{eff}}{V_0} \frac{\gamma_m^2}{\gamma_{spp}}$, which in the nearfield assumes a value $\kappa^* = 1940\ ps^{-1}$ corresponding to a quality factor $Q = \frac{\omega_{spp}}{\kappa^*}$ (= $1.1 \times 10^6$) (Table S1). According to the equation $\gamma'_m = \gamma_m\ Q/V$, the multilayer volume mode and the quality factor modify the radiative rate of the molecular dipole. Besides, in the updated optomechanics, the cavity/dye has a nearfield radiation rate schematized as $\gamma_f = \gamma'_m \frac{\kappa^*}{\kappa^* + \gamma_{ohm}}$ and, altogether, one can conclude that radiative rate outperforms the cavity decay of a factor $\frac{\gamma_f}{\kappa^*} = 4.4$.

**Table S1**. Factors and values of the system

| Variable | $\gamma_m$ | $\omega_m$ | V | $\gamma_f$ | $\gamma_{ohm}$ | $\gamma_{spp}$ | $\kappa *$ | Q | $\gamma'_m$ | $\omega_{spp}$ |
|---|---|---|---|---|---|---|---|---|---|---|
| Value | $285\ ps^{-1}$ | $5.5 \times 10^8\ MHz$ | $8 \times 10^7$ | $8461\ ps^{-1}$ | $2 \times 10^6 ps^{-1}$ | $3 \times 10^9\ ps^{-1}$ | $1940\ ps^{-1}$ | $1.1 \times 10^6$ | $8.7 \times 10^6 ps^{-1}$ | $2.2 \times 10^9\ MHz$ |

### §S3. Enantioselectivity

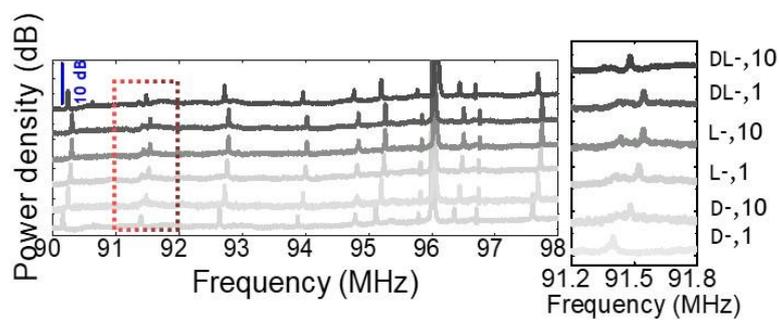

**Figure S1.** Transmission spectrum (D)-penicillamine and (L)- penicillamine, and racemic solution DL; 1, 10 stands for 1 mM and 10 mM. The range of frequency in the dashed rectangles is shown at the right of the panel.

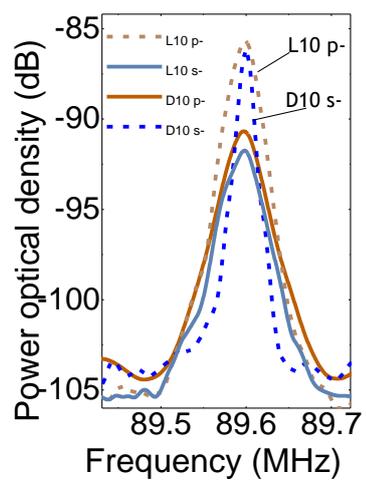

**Figure S2.** Discriminative peak of the multilayer (no rhodamine).

14